\documentstyle[twoside,psfig]{article}

\catcode`\@=11
\long\def\@makefntext#1{
\protect\noindent \hbox to 3.2pt {\hskip-.9pt
$^{{\eightrm\@thefnmark}}$\hfil}#1\hfill}               

\def\@makefnmark{\hbox to 0pt{$^{\@thefnmark}$\hss}}    

\def\ps@myheadings{\let\@mkboth\@gobbletwo
\def\@oddhead{\hbox{}
\rightmark\hfil\eightrm\thepage}
\def\@oddfoot{}\def\@evenhead{\eightrm\thepage\hfil
\leftmark\hbox{}}\def\@evenfoot{}
\def\sectionmark##1{}\def\subsectionmark##1{}}

\oddsidemargin=\evensidemargin
\newcommand{\beq}{\begin{equation}}
\newcommand{\eeq}{\end{equation}}

\newcounter{sectionc}\newcounter{subsectionc}\newcounter{subsubsectionc}
\renewcommand{\section}[1] {\vspace{12pt}\addtocounter{sectionc}{1}
\setcounter{subsectionc}{0}\setcounter{subsubsectionc}{0}\noindent
        {\tenbf\thesectionc. #1}\par\vspace{5pt}}
\renewcommand{\subsection}[1] {\vspace{12pt}\addtocounter{subsectionc}{1}
        \setcounter{subsubsectionc}{0}\noindent
        {\bf\thesectionc.\thesubsectionc. {\kern1pt \bfit #1}}\par\vspace{5pt}}
\renewcommand{\subsubsection}[1] {\vspace{12pt}\addtocounter{subsubsectionc}{1}
        \noindent{\tenrm\thesectionc.\thesubsectionc.\thesubsubsectionc.
        {\kern1pt \tenit #1}}\par\vspace{5pt}}
\newcommand{\nonumsection}[1] {\vspace{12pt}\noindent{\tenbf #1}
        \par\vspace{5pt}}

\newcounter{appendixc}
\newcounter{subappendixc}[appendixc]
\newcounter{subsubappendixc}[subappendixc]
\renewcommand{\thesubappendixc}{\Alph{appendixc}.\arabic{subappendixc}}
\renewcommand{\thesubsubappendixc}
        {\Alph{appendixc}.\arabic{subappendixc}.\arabic{subsubappendixc}}

\renewcommand{\appendix}[1] {\vspace{12pt}
        \refstepcounter{appendixc}
        \setcounter{figure}{0}
        \setcounter{table}{0}
        \setcounter{lemma}{0}
        \setcounter{theorem}{0}
        \setcounter{corollary}{0}
        \setcounter{definition}{0}
        \setcounter{equation}{0}
        \renewcommand{\thefigure}{\Alph{appendixc}.\arabic{figure}}
        \renewcommand{\thetable}{\Alph{appendixc}.\arabic{table}}
        \renewcommand{\theappendixc}{\Alph{appendixc}}
        \renewcommand{\thelemma}{\Alph{appendixc}.\arabic{lemma}}
        \renewcommand{\thetheorem}{\Alph{appendixc}.\arabic{theorem}}
        \renewcommand{\thedefinition}{\Alph{appendixc}.\arabic{definition}}
        \renewcommand{\thecorollary}{\Alph{appendixc}.\arabic{corollary}}
        \renewcommand{\theequation}{\Alph{appendixc}.\arabic{equation}}
        \noindent{\tenbf Appendix \theappendixc #1}\par\vspace{5pt}}
\newcommand{\subappendix}[1] {\vspace{12pt}
        \refstepcounter{subappendixc}
        \noindent{\bf Appendix \thesubappendixc. {\kern1pt \bfit #1}}
        \par\vspace{5pt}}
\newcommand{\subsubappendix}[1] {\vspace{12pt}
        \refstepcounter{subsubappendixc}
        \noindent{\rm Appendix \thesubsubappendixc. {\kern1pt \tenit #1}}
        \par\vspace{5pt}}

\newcommand{\textlineskip}{\baselineskip=13pt}
\newcommand{\smalllineskip}{\baselineskip=10pt}

\def\eightcirc{
\begin{picture}(0,0)
\put(4.4,1.8){\circle{6.5}}
\end{picture}}
\def\eightcopyright{\eightcirc\kern2.7pt\hbox{\eightrm c}}

\def\abstracts#1#2#3{{
        \centering{\begin{minipage}{4.5in}\footnotesize\baselineskip=10pt
        \parindent=0pt #1\par 
        \parindent=15pt #2\par
        \parindent=15pt #3
        \end{minipage}}\par}} 

\newcommand{\bibit}{\nineit}

\renewenvironment{thebibliography}[1]
        {\frenchspacing
         \ninerm\baselineskip=11pt
         \begin{list}{\arabic{enumi}.}
        {\usecounter{enumi}\setlength{\parsep}{0pt}     
         \setlength{\leftmargin 12.7pt}{\rightmargin 0pt} 
         \setlength{\itemsep}{0pt} \settowidth
        {\labelwidth}{#1.}\sloppy}}{\end{list}}

\newcounter{itemlistc}
\newcounter{romanlistc}
\newcounter{alphlistc}
\newcounter{arabiclistc}

\newcommand{\fcaption}[1]{
        \refstepcounter{figure}
        \setbox\@tempboxa = \hbox{\footnotesize Fig.~\thefigure. #1}
        \ifdim \wd\@tempboxa > 5in
           {\begin{center}
        \parbox{5in}{\footnotesize\smalllineskip Fig.~\thefigure. #1}
            \end{center}}
        \else
             {\begin{center}
             {\footnotesize Fig.~\thefigure. #1}
              \end{center}}
        \fi}

\newcommand{\tcaption}[1]{
        \refstepcounter{table}
        \setbox\@tempboxa = \hbox{\footnotesize Table~\thetable. #1}
        \ifdim \wd\@tempboxa > 5in
           {\begin{center}
        \parbox{5in}{\footnotesize\smalllineskip Table~\thetable. #1}
            \end{center}}
        \else
             {\begin{center}
             {\footnotesize Table~\thetable. #1}
              \end{center}}
        \fi}

\def\@citex[#1]#2{\if@filesw\immediate\write\@auxout
        {\string\citation{#2}}\fi
\def\@citea{}\@cite{\@for\@citeb:=#2\do
        {\@citea\def\@citea{,}\@ifundefined
        {b@\@citeb}{{\bf ?}\@warning
        {Citation `\@citeb' on page \thepage \space undefined}}
        {\csname b@\@citeb\endcsname}}}{#1}}

\newif\if@cghi
\def\cite{\@cghitrue\@ifnextchar [{\@tempswatrue
        \@citex}{\@tempswafalse\@citex[]}}
\def\citelow{\@cghifalse\@ifnextchar [{\@tempswatrue
        \@citex}{\@tempswafalse\@citex[]}}
\def\@cite#1#2{{$\null^{#1}$\if@tempswa\typeout
        {IJCGA warning: optional citation argument 
        ignored: `#2'} \fi}}

\def\pmb#1{\setbox0=\hbox{#1}
        \kern-.025em\copy0\kern-\wd0
        \kern.05em\copy0\kern-\wd0
        \kern-.025em\raise.0433em\box0}

\def\fnt#1#2{\footnotetext{\kern-.3em
        {$^{\mbox{\scriptsize #1}}$}{#2}}}

\def\ps@myheadings{%
    \let\@oddfoot\@empty\let\@evenfoot\@empty
    \def\@evenhead{\slshape\leftmark\hfil}
    \def\@oddhead{\hfil{\slshape\rightmark}}
    \let\@mkboth\@gobbletwo
    \let\sectionmark\@gobble
    \let\subsectionmark\@gobble
    }
\font\tenrm=cmr10
\font\tenit=cmti10 
\font\tenbf=cmbx10
\font\bfit=cmbxti10 at 10pt
\font\ninerm=cmr9
\font\nineit=cmti9

\font\eightrm=cmr8

\def\qed{\hbox{${\vcenter{\vbox{                        
   \hrule height 0.4pt\hbox{\vrule width 0.4pt height 6pt
   \kern5pt\vrule width 0.4pt}\hrule height 0.4pt}}}$}}

\begin{document}
\setlength{\textheight}{7.7truein}  

\thispagestyle{empty}

\normalsize\textlineskip

\setcounter{page}{1}

\vspace*{0.88truein}

\centerline{\bf Mesonic Content of the Nucleon and the Roper Resonance\footnote{Supported in part by Forschungszentrum FZ J\"ulich (COSY)}}\baselineskip=13pt
\vspace*{0.4truein}
\centerline{\footnotesize M.Dillig \footnote{Email: mdillig@theorie3.physik.uni-erlangen.de} and M.Schott}
\baselineskip=12pt
\centerline{\footnotesize\it Institute for Theoretical Physics III\footnote{preprint FAU-TP3-06/Nr. 08}} \, 
\centerline{\footnotesize\it University of
Erlangen-N\"urnberg}
\centerline{\footnotesize\it Staudtstr. 7, Erlangen, D-91058, Germany}
\vspace*{12pt}

\vspace*{0.23truein} \abstracts{We investigate colorless (mesonic)
  $3q-q\bar q$ components in the nucleon and the Roper resonance
  $N^*(1440)$. Starting from constituent quarks and gluons we estimate
  the excitation of $q \bar q$ pairs in a gluon exchange model with the
  strong coupling constant extracted from a comparison with a
  non-perturbative resonating group calculation of the $\pi, \eta,
  \rho$ and $\omega$ content of the nucleon. Applying the same model
  to the Roper resonance, we find as the most striking result a very
  strong scalar-isoscalar $\sigma$-content of the $N^*(1440)$ with a
  strength of the $3q-q\bar q$ configurations comparable to the $3q$
  component itself.}{}{}
\vspace*{10pt}
\vspace*{2pt}

%
%
%
%
\baselineskip=13pt              
\normalsize                     

\vspace*{-0.5pt}
\noindent

\bigskip
\vspace*{12pt}

Among the excited states of the nucleon, i.e. the baryon resonances in
the continuum, the Roper resonance $N^*(1440)$ as the first excited
state with the quantum numbers of the nucleon $(J^\pi,T)=
(\frac{1}{2}^+, \frac{1}{2})$, plays a particular role. On the one side, parameterizing conventionally the Roper
resonance as a system of three constituent quarks, most quark models
predict the position of the Roper resonance as generally several
hundred MeV above its experimental value [1-3]. On the other side, the
$N^*(1440)$ seems to dominate the dynamics of coherent near threshold
$2\pi$ production, both in $pp\rightarrow pp \pi \pi$ [4-7] and $\pi A
\rightarrow \pi \pi A$ collisions [8,9], through the coupling to a
virtual $\sigma$-meson (which subsequently then decays into two
pions). Thus it is hoped that from the study of the Roper-induced
dynamics ultimately shed light on the $\sigma$-degree of freedom -
which is generally phrased as a $2\pi$ resonance with a mass and width
of $m _\sigma \sim \Gamma _\sigma \sim 550 MeV$ [10] - both in the NN and
in nuclear systems, where the $\sigma$-meson provides the medium range
attraction; in addition, it might provide information of a partial
restoration of chiral symmetry in dense nuclear matter through a
strong reduction of the effective $\sigma$-meson mass down to twice
the pion mass [11,12].\\
   
One source of the problem in describing the Roper resonance - and,
in general, all unstable baryon resonances - in quark models, is
the restriction of their quark content to three constituent quarks.
Already the experimentally verified strong decay of the Roper
resonance with a width of $\Gamma \sim 240 - 450 MeV$ into $N\pi$ and
$N\pi \pi$ final states [10], signals a strong next order $4q - \bar q$
component in the Fock expansion
\begin{equation}|N^*\rangle  = (1-\alpha) |3 q\rangle  + \alpha |3 q (q \bar q)\rangle \end{equation}
where in color-singlet channels the $(q \bar q)$ content can be
classified along experimentally known mesons [13-17].\\

In this note we investigate the mesonic structure of the Roper
resonance, with the emphasis in estimating its scalar
$\sigma$-content. In more detail, we would like to pursue the
following route. Starting from a Hamiltonian which includes the one
gluon exchange (OGE), harmonic confinement and in addition, $q\bar q$
pair creation and annihilation, the mesonic content of the component
$3q-(q\bar q)$ of the nucleon is calculated in a non-perturbative
resonating group calculation for $\pi, \eta, \rho$ and $\omega$ from
minimizing [18,19]
\begin{equation}\delta (\frac{\langle \Psi|H-E|\Psi\rangle }{\langle \Psi|\Psi\rangle })=0\end{equation}
We then simulate this result in a perturbative one-loop calculation,
with $q\bar q$ pair creation induced by constituent gluon
exchange [18-23]. With the strength of the 
effective $qqg$ vertex given, we
then deduce the $\sigma$-content of the nucleon and extent the model
into the continuum, to estimate the mesonic content of the Roper
resonance in $N^*(1440)\rightarrow (N,\lambda)$ transitions ($\lambda$ denotes 
the corresponding meson). In comparing meson creation by
 one gluon exchange with the
 $^3P_0$ vacuum pair creation model [24,25], the basic difference
 to eq. (14) below
is its simplified spin structure $\sim \vec\sigma \vec q$. \\

We briefly sketch the main lines of our calculation, deferring,
however, technical details to a forthcoming comprehensive publication.
The basic formulae for calculating the admixture of a nucleon and a
meson $\lambda$ in the nucleon itself or the Roper resonance is given
as
\begin{equation}\langle B|B\rangle  = (1-\alpha )^2 \langle 3q|3q\rangle  + \alpha ^2 \langle 3q(q \bar q)|3q (q \bar q)\rangle \end{equation}
where the admixture probability $\alpha ^2$ is given in the one-loop
approximation as [26,27]
\begin{equation}\alpha ^2 = \textbf{P} \int 
\frac{|M _{B\rightarrow N \lambda} (\vec q)|^2 d\vec q}{2 
\omega _\lambda (\vec q) (\omega _\lambda (\vec q) +E _N (\vec q) - 
M _B)^2} d(\vec q) \end{equation}
where the numerator contains the $B\rightarrow N\lambda$ transition
amplitude $M _{B\rightarrow N \lambda} (\vec q)$, while $M _B$
represents the $N$ or $N^*$ mass, respectively (we assume the initial
baryon to be at rest); $\omega _\lambda (\vec q)$ and $E _N (\vec q)$
are the total energies of the meson and the nucleon; The symbol
\textbf{P} denotes the principal value of the integral: for $m
_\lambda + M > M _B$ it yields a standard integral without
singularity. Note that for $m _\lambda + M \le M _B$ the principal
value is conveniently evaluated from the relation
\begin{equation} \textbf{P} \int _0 \frac{f(q) dq}{q^2 - q _0^2} = 
\int_0^\infty \frac{f(q) - f(q _0)}{q^2 - q_0^2} dq + 
\frac{i \pi}{q _0} f(q_0)\end{equation}
where $q _0$ denotes the position of the singularity.\\

The dynamics of the admixture is contained in the transition amplitude
$M _\lambda (\vec q)$ given as (in an obvious notation)
\begin{equation}M _\lambda (\vec q) = \langle N(3q; \vec q) 
\lambda (q \bar q, \vec q) | V _{B\rightarrow N \lambda} (\vec q) | 
N_B (3q)\rangle \end{equation}
More explicitly $M _\lambda (\vec q)$ factorizes into a $\vec q$-dependent
piece times spin, flavor and color coefficients, i.e.
\begin{equation}M _\lambda (\vec q) = T _\lambda (\vec q) 
\lambda _{S} \lambda _F \lambda _C\end{equation}
for all quarks in relative s - states; in case of the $\sigma$ meson
the orbital angular momentum l = 1 is coupled with the $q\bar q$ (see eq. (13) 
below).\\

For an economical evaluation we introduce two additional
simplifications: we represent the relative motion between the $3q$ and
$q \bar q$ clusters by a plane wave (through a full resonating group
calculation yields a self consistent relative wave function between the clusters) and we
model the nucleon and the $N^*$ as a quark - (scalar - isoscalar)
diquark system in a relative $0s$ and $1s$ state
for the nucleon and the Roper resonance, respectively; both
assumptions simplify in particular anti-symmetrization among the
quarks and the evaluation of the color matrix elements significantly
). This diquark 
picture reflects the strong qq correlations in the scalar - isoscalar channel
and is supported by numerous investigations, such as on the lattice [28,29], Dyson - Schwinger studies [30,31] and 
from the non relativistic quark model [32,33].
This
yields then for the complete wave function
\begin{equation}\Phi _\lambda |(q \bar q)\rangle  = 
\Phi _\lambda (\rho) [\frac{1}{2}\frac{\bar 1}{2}]^s]_\lambda
[\frac{1}{2}\frac{1}{2}]_\lambda [(10)(01)]_\lambda^{00}\end{equation}
for the $\pi, \eta, \rho$ and $\omega$ mesons and correspondingly
\begin{equation}|(3q)_B\rangle  = \Phi _B (r) 
[\frac{1}{2} 0]^{\frac{1}{2}m} [\frac{1}{2}0]^{\frac{1}{2}M} 
[(10)(01)]^{00}\end{equation}
for the baryons. Above, the bracket $[ab]^{JM}$ denotes appropriate
spin and isospin coupling; the color wave function, which in our
diquark picture becomes identical for all mesons and baryons, is
easily reduced from the Elliot notation [34] to
\begin{equation} [(10)(01)]^{00} = \frac{\delta _{ab}}{\sqrt{3}}\end{equation}
with a and b referring to the q, diquark or $q,\bar q$ color content,
respectively.  The radial dependence is then parameterized in a
translational invariant normalized Gaussian form as
\begin{equation}\Phi (r) = \sum _i N _i e^{-\frac{r^2}{2 b _i^2}}\end{equation}
where for the mesons $\pi, \eta, \rho$ and $\omega$ the sum over i
includes only a single Gaussian form, which is related via
\begin{equation}b _i = \frac{2}{\sqrt{3}} r _{r.m.s.}\end{equation}
to the mean $rms$-radius of the mesons, whereas for the parameterization of the baryons 3 Gaussians with
 different width parameters are included ([19], compare Table 1). Only for the
$\sigma$-meson with positive parity the radial structure is different: the
$q\bar q$ pair is in a relative $l=1$ orbit [35], yielding
\begin{equation}\Phi _\sigma (\vec r) = N _\sigma e^{-\frac{r^2}{2 
b_\sigma
^2}} [Y _1 (\hat {\vec r}) [\frac{1}{2}\frac{\bar 1}{2}]^1]^{00}
\end{equation}

Finally we specify the $B\rightarrow N \lambda$ transition operator.
Here we focus on an effective one gluon exchange for the excitation of the
$q\bar q$ pair. This yields in momentum space [18,19]
obtain in momentum space (Fig. 1)
\begin{equation} V _{B\rightarrow N \lambda}(\vec q) = 4\pi\frac{\alpha_s}{m_q}
\frac{\lambda_i \lambda_j}{4} 
\frac{(\vec\sigma_i \times  \vec\sigma_j)\vec q -2 \vec\sigma_j \vec p_i}
{\vec q^2 + m _g^2} \end{equation}
(with quark and gluon mass $m _q$ and $m _g$, respectively; $\vec p_i$ is the momentum of the initial quark, from which the $q\bar q$ pair is exited), which can
be cast for $m _g^2 \gg \vec q^2$ into a Gaussian form in coordinate
space
\begin{equation} V _{B\rightarrow N \lambda}(\vec r) = - i (\frac{\alpha_s}
{8\sqrt \pi} \frac{m _g^3}{m _q} \frac{\lambda_i \lambda_j}{4}
((\vec \sigma_i \times \vec \sigma_j) \times \vec r_{ij} 
e^{- \frac{m _g ^2 r_{ij}^2}{4}}
+ i \, 4e^{- \frac{m _g ^2 r_{ij}^2}{4}} \vec \sigma_j \vec \nabla_i))
\end{equation}
The strong coupling constant $\alpha_s \sim 2$ is kept as a parameter, 
which is 
adjusted to reproduce in our one-loop approximation the admixture
probabilities obtained from a corresponding non-perturbative resonating group
calculation later on.\\

\begin{figure}[htbp] 
  \vspace*{13pt}
  \centerline{\framebox{\psfig{width=5cm,file=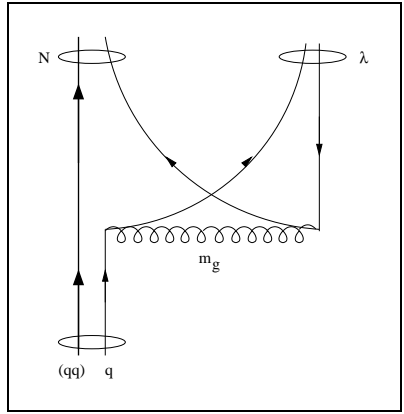}}}    
  \vspace{0cm}
  \hspace{2.5cm}\parbox{7.5cm}{
  \caption{Schematical representation of a mesonic $q \bar q$ excitation in an
(effective) one-gluon exchange model.}}                                                                                                              
\end{figure}

The formalism, developed in a Gaussian representation given above,
allows a very transparent evaluation of the admixture probability: the

complete transition amplitude, involving multiple Gaussian integrals,
can be evaluated analytically; only the evaluation of the principal
value in eq. 5 has to be performed numerically.\\

For the final result we specify our model parameters as follows: we use $m _q =
m _g / 2 = 330 MeV$ for the constituent quark and gluon masses [36,37]; the
Gaussian parameters for the participating particles are given in Table 1 [19]

\begin{table}[htbp]
\label{TabelleHoffmann}
\centerline{\footnotesize\smalllineskip
\begin{tabular}{|l|c|c|c|c|c|c|}
\hline
{} & $\beta _1$ & $\beta _2$ & $\beta _3$ & $b _1$ & $b_2$ & $b _3$\\
\hline
$N$ & 1.1712 & 1.303 & 0.021 & 0.440 & 0.659 & 0.911 \\
\hline
$N^*$ & 4.625 & 0.418 & -0.822 & 0.440 & 0.659 & 0.911 \\
\hline
\hline
{} & $\pi$ & $\eta$ & $\sigma$ & $\rho$ & $\omega$ & {}\\
\hline
$b_\lambda$ & 0.5 & 0.5 & 0.7 & 0.75 & 0.6 & {}\\
\hline
\end{tabular}}
\caption{Parameters for the $N,N^*$ and mesons $\lambda$ (given in [fm])}
\end{table}

The effective coupling strength $\alpha _s$, if
fitted to the probabilities of RG calculation in the nucleon with
\begin{equation} P _\pi = 0.045,  P _\eta = 0.005; P _\rho = 0.03, 
P _\omega = 0.004\end{equation}
leads qualitatively to $\alpha _s \cong 1.8$. Then, as a first result,
we estimate the $\sigma$-content in the nucleon as $P_\sigma = 0.025$
(Fig.2). In going then over to the mesonic admixtures in the Roper
resonance, the admixture probabilities and their relative importance
are changed in part dramatically. For the $\pi, \rho$ and $\omega$ the
probabilities remain qualitatively close to the content in the
nucleon. This is not surprising, as for the pion with $m _\pi + M <
M^*$, the singular piece in the principal value is strongly canceled,
while the $\rho$ and $\omega$ admixture are still fairly small due to
$m _{\rho,\omega} + M - M^* \sim 280 MeV$. For the $\sigma$ and the
$\eta$, however, the situation is quite different: here, from $m
_{\sigma, \eta} + M - M^* \sim 50 MeV$, already the squared energy
denominators enhance the admixture significantly;
explicitly we find an admixture probability of typically $50$ percent
for the $\sigma$ meson and of $10$ percent for the $\eta$-meson,
respectively; these characteristic values are fairly stable against a
moderate variation of the Gaussian model parameters: the
maximal sensitivity of the result is found with respect to the root
mean square radii of the mesons (Fig.3).\\

The various conclusions to be drawn from our estimate, are evident:

\begin{itemize}
  
\item the scalar-isoscalar $\sigma$-degree of freedom has a strong
  component in the Roper resonance [38]; consequently, as the
  $\sigma$-meson strongly couples to the $\pi \pi$ channel, the
  $\sigma$ dominates hadron or photon induced double pion production
  in the $N^* (1440)$ regime [39,40];
  
\item the $\eta$-meson also strongly couples to the Roper resonance
  (in fact, in former tables of particle properties the decay $N^*
  (1440)\rightarrow N + \eta$ was listed with a fairly large
  strength). An important consequence might be, that the $\eta NN^*$
  coupling strongly contributes to the near threshold production of
  the $\eta$-meson [41], where the threshold is only about $50MeV$ away;

\item the large $q \bar q$ components in the Roper resonance suggest
  as a universal feature of baryon resonances in the continuum, that
  any quantitative treatment of $N^*$'s requires the inclusion of, at
  least, the next - to leading $3q(q\bar q)$ component to the leading
  $3q$ piece in a corresponding Fock-expansion. As an immediate
  consequence we expect a significant shift of the (real part of the)
  $N^*$-masses of the order of the resonance widths, which would spoil
  the in part quantitative result obtained for baryon spectroscopy in
  various quark models.

\end{itemize}

\begin{figure}[htbp] 
\vspace*{13pt}
\framebox{\centerline{\psfig{height=7.5cm,file=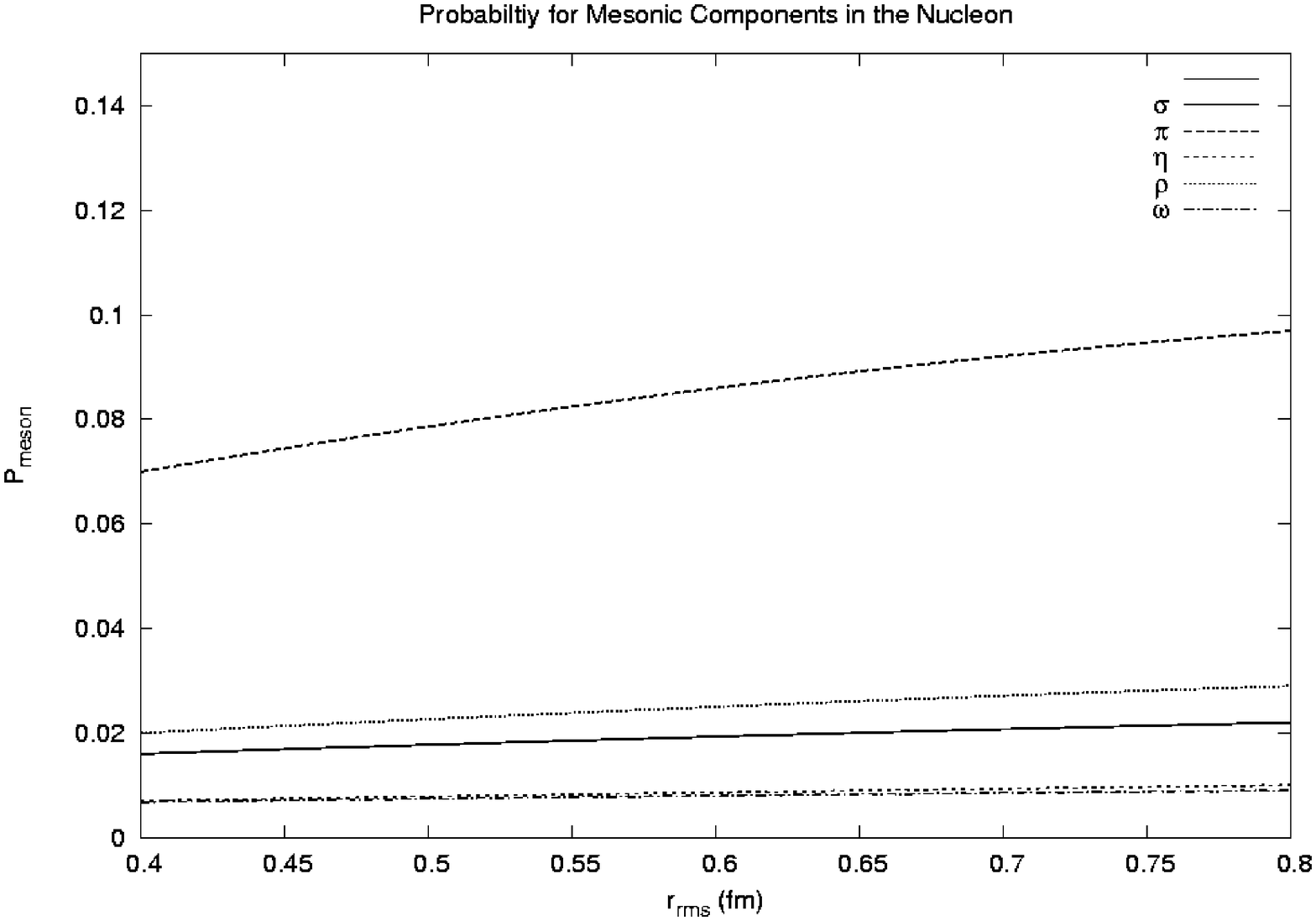}}} 
\vspace*{13pt}
\vspace{-0.5cm}
\caption{Admixture probability of various meson contributions (with masses below 1GeV) to the
nucleon as a function of the mesonic root-mean-square radius $r _{rms}$}
\end{figure}

\begin{figure}[htbp] 
\vspace*{13pt}
\framebox{\centerline{\psfig{height=7.5cm,file=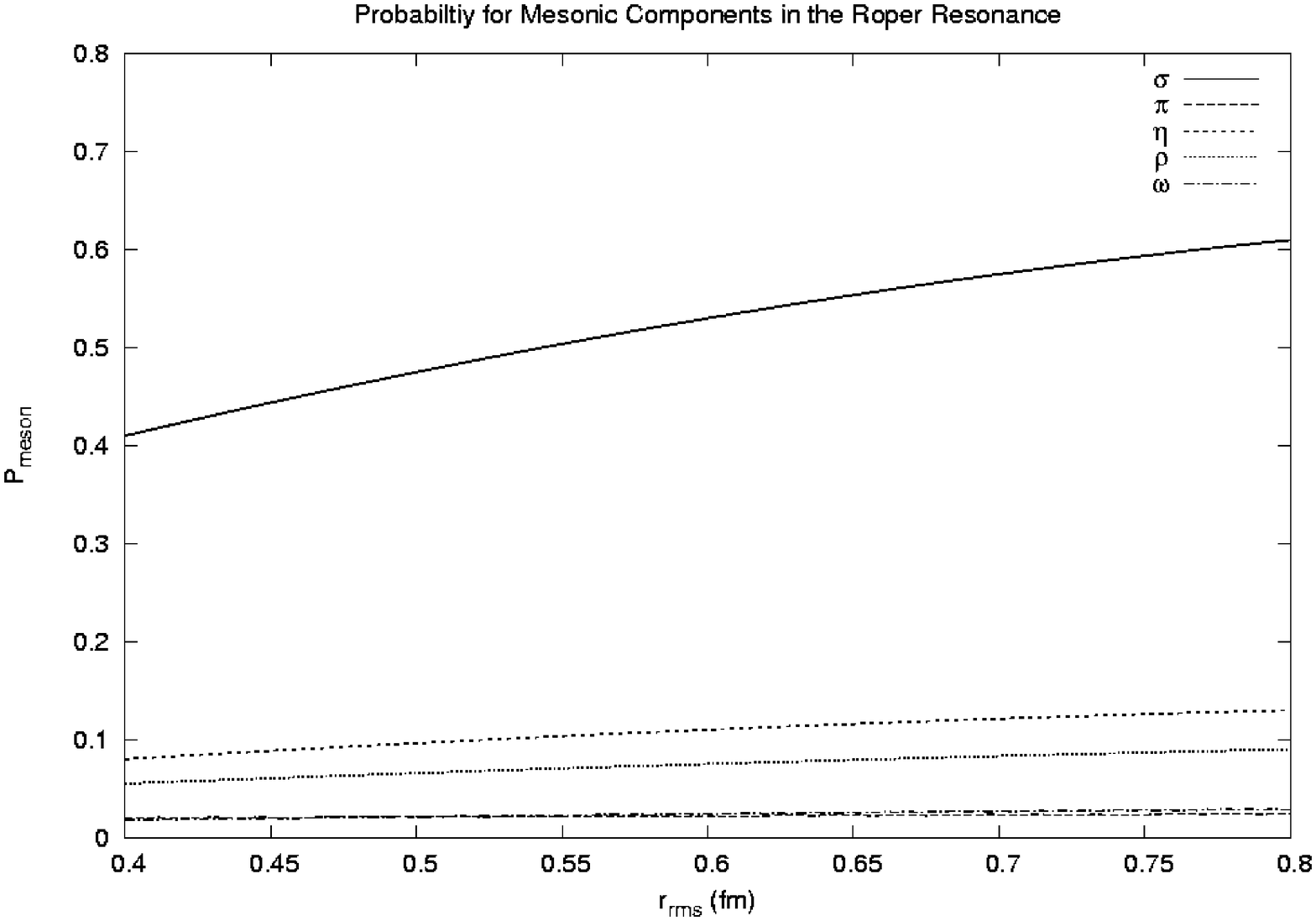}}} 
\vspace*{13pt}
\vspace{-0.5cm}
\caption{As Fig.2, however, for the Roper resonance $N^*(1440)$}
\end{figure}

Of course, for firmer conclusions, our model estimate in this note has
to be improved, such as to include explicitly the $3q$-component of
baryons (instead of a quark-diquark representation) or a consistent
handling of the relative $N \lambda$- cluster wave function.\\
 
Beyond that the representation
of the $\sigma$-meson as a correlated $\pi \pi$ system [42] has to be
investigated in detail. Even though this picture of a $\sigma$-meson as a $(q
\bar q)(q \bar q)$ molecule is very different from viewing the
$\sigma$ as a dominant $q\bar q$ system, we still expect simply from
parity reasons, that $\sigma$ behaves quite different from the other
mesons with masses below $1 GeV$.\\ 

Just to present the main argument. Assuming for simplicity identical root mean square radii for all mesons considered,
leads to an identical product of the radial wave functions at the $N^*N\lambda$
vertices; thus the particular role of the $\sigma$ meson is related to its orbital - spin structure $[Y_1 [\frac{1}{2}\frac{1}{2}]^1]^{00}$ in contrast
to $[Y_0 [\frac{1}{2}\frac{1}{2}]^S]^{SM}$, with S = 0,1,
 for the $\pi,\eta$ and
$\rho,\omega$, respectively. Consequently, recoupling the above invariants with the $q\bar q$ creation operator $[Y_1\Sigma]^{00}$ yields
the production operator itself (with the angular momentum function $Y_1$)
for $\pi, \eta, \rho, \omega$, whereas for the $\sigma$ meson 
\begin{equation}
[Y_1[\frac{1}{2}\frac{1}{2}]^1]^0 [Y_1 \Sigma]^0]^{00} = 
\sum_{L=0,2} {\hat L} [[Y_1 Y_1]^L[\frac{1}{2}\frac{1}{2}]^1]^L]^{00}
\end{equation}
it results in a very different spin - orbit structure. This very difference with L=0,2, which is reflected in the large $\sigma$ content of the Roper resonance, is preserved for a $\pi\pi$ structure of the $\sigma$: recoupling
$\sigma = [[\frac{1}{2}\frac{1}{2}]^0[\frac{1}{2}\frac{1}{2}]^0]^{00}$ with 
twice the production operator yields immediately
\begin{equation}
[[Y_1\Sigma^{\prime}]^0  [Y_1 \Sigma]^0]^{00} = 
\sum_{L=0,2} \hat {L} [[Y_1 Y_1]^L [\Sigma^{\prime}\Sigma]^L]^{00}
\end{equation}
i.e. the same structure as for the representation of $\sigma$ as a $q\bar q$ 
pair in eq. (15).\\

To substantiate our findings and conjectures and
to extend the model predictions in a systematic way to other baryon
resonances, detailed calculations are in progress; furthermore, more insight will be gained from detailed information on the internal structure of baryon resonances as under intensive study at modern hadron (COSY) [43] and electron accelerators (MAMI, ELSA,CEBAF)[44],[45],[46].
\bigskip

The authors thank H.M.Hofmann for fruitful discussions.

\nonumsection{References}

\end{document}